\documentclass{aa}
\usepackage{graphicx}
\usepackage{float}
\usepackage{hyperref}
\usepackage{placeins}

\begin{document}
\title{POLAR-II: modeling star formation history of galaxies on the 21-cm signal from Epoch of Reionization}
\titlerunning{POLAR-II}


\author{Qing-Bo Ma\inst{1,2}\thanks{\email{maqb@gznu.edu.cn}}
\and Raghunath Ghara\inst{3}
\and Benedetta Ciardi\inst{4}
\and Anshuman Acharya\inst{4,5}
\and Bin Yue\inst{6,7}
\and Ilian T. Iliev\inst{8}
\and Léon V. E. Koopmans\inst{9}
\and Garrelt Mellema\inst{10}
\and Saleem Zaroubi\inst{9,11}
\\
     }
\authorrunning{Q. Ma et al.}

\institute{School of Physics and Electronic Science, Guizhou Normal University, Guiyang 550001, PR China
\and Guizhou Provincial Key Laboratory of Radio Astronomy and Data Processing, Guizhou Normal University, Guiyang 550001, PR China
\and Department of Physical Sciences, Indian Institute of Science Education and Research Kolkata, Mohanpur, WB 741 246, India
\and Max-Planck Institute f\"ur Astrophysics, Karl-Schwarzschild-Stra\ss e 1, 85748 Garching bei M\"unchen, Germany
\and Berkeley Center for Cosmological Physics, Department of Physics, University of California, Berkeley, CA 94720, USA
\and National Astronomical Observatories, Chinese Academy of Sciences, 20A Datun Road, Chaoyang District, Beijing 100101, China
\and State Key Laboratory of Radio Astronomy and Technology, Beijing 100101, China
\and Astronomy Centre, Department of Physics and Astronomy, University of Sussex, Falmer, Brighton BN19QH, UK
\and Kapteyn Astronomical Institute, University of Groningen, P.O. Box 800, NL-9700AV Groningen, the Netherlands
\and The Oskar Klein Centre, Department of Astronomy, Stockholm University, AlbaNova, SE-10691 Stockholm, Sweden
\and ARCO (Astrophysics Research Center), Department of Natural Sciences, The Open University of Israel, 1 University Road, PO Box 808, Ra’anana 4353701, Israel
\\}

\date{Received XX XX, 20XX}

\abstract
{Galaxies may suffer some starburst and quenched periods in their history due to e.g. galaxy mergers and feedback.
However, semi-numerical simulations of the Epoch of Reionization (EoR) typically do not accurately model the effects of the star formation history (SFH) of galaxies.}
{{Keeping the same total ionizing photon budget from galaxies,} we investigate how the ionization and heating of the Intergalactic Medium (IGM), as well as the associated 21-cm signal during the EoR, depends on the variations in the modeling of the SFH of galaxies.}
{We adopt the {\sc Jiutian-300} {\it N}-body dark matter simulation and the semi-analytic model {\sc L-Galaxies 2020} to model galaxy formation.
Using the galaxy catalog from {\sc L-Galaxies 2020} as input, we post-process the {\sc Jiutian-300} density field with the one-dimensional radiative transfer code {\sc Grizzly} to model the reionization process and the 21-cm signal.}
{We find that the ionized regions produced by galaxies with a SFH derived from {\sc L-Galaxies 2020} are slightly larger and warmer than the ones obtained with a constant SFR.
For a fixed stellar mass, galaxies produce smaller ionized regions with increasing stellar mass weighted stellar age $\tau_{\rm age}$.
This results in a different topology and timing of the IGM ionization and heating obtained from {\sc Grizzly}.}
{The SFH of galaxies is highly dependent on $\tau_{\rm age}$ and redshift.
Different models of the galactic SFH affect the gas heating and ionizing processes during the EoR, and as a consequence also the 21-cm global signal and power spectrum.}


\maketitle
\nolinenumbers

\section{Introduction}
\label{sec:intro}
The epoch of reionization (EoR) is a major phase transition of the Universe from fully neutral and cold to highly ionized and hot \citep{Gnedin2022LRCA}, which happens after the formation of the first stars and galaxies and ends with the full ionization of the neutral Intergalactic Medium (IGM).
Indirect observations of the physical state of the IGM, e.g. the Gunn–Peterson absorption trough in the high-$z$ QSO spectra \citep{Fan2006, Bosman2022} and the Thomson scattering optical depth measured by Cosmic Microwave Background (CMB) experiments  \cite[e.g. Planck satellite][]{Planck2020A&A}, suggest that the EoR ends at $z \gtrsim 5$.

Since there is abundant neutral IGM during the EoR, the 21-cm hyperfine line of neutral hydrogen is expected to be an ideal probe of the high-$z$ Universe \citep{Furlanetto2006}.
The project Experiment to Detect the Global EoR Signature (EDGES experiment)\footnote{https://loco.lab.asu.edu/edges/} reported an absorption profile of 21-cm signal at $78 \,\rm MHz$ \cite[i.e. $z \sim 17$,][]{Bowman2018}, which has been interpreted as associated to the formation of the first sources in the Universe.
This result is strongly debated \cite[e.g.][]{Hills2018, Singh2022} and the absorption feature has not been found by the experiment SARAS 3 \citep{Bevins2022}.
Radio array telescopes such as the low-frequency array (LOFAR)\footnote{https://www.astron.nl/telescopes/lofar/}, the New Extension in Nançay Upgrading LOFAR (NenuFAR)\footnote{https://nenufar.obs-nancay.fr/}, the hydrogen Epoch of Reionization Array (HERA)\footnote{https://reionization.org}, and the Murchison Widefield Array (MWA)\footnote{https://www.mwatelescope.org/}, have presented results on the upper limit of 21-cm power spectra \cite[e.g.][]{Mertens2025A&A, Munshi2025MNRAS, Abdurashidova2022, Trott2025arXiv}, which have already been used to rule out some EoR models \citep{Ghara2025A&A, Mondal2020}.
The next generation radio telescope Square Kilometre Array (SKA)\footnote{https://www.skao.int/en} will provide more details on the history and topology of the EoR \citep{Koopmans2015}.

Observations of the IGM with 21-cm experiments are complemented by recent data on high-$z$ galaxies.
Due to its high sensitivity and wavelength coverage, in the last few years the James Webb Space Telescope (JWST) has provided data on low-mass and faint galaxies at high-$z$.
These include three galaxies with spectroscopically confirmed redshifts of $z\sim 14$ \citep{Carniani2024Natur,Naidu2025arXiv},
the rest-frame UV luminosity functions (UVLF) of galaxies up to $z \sim 16$ \citep{Harikane2023ApJS, Donnan2024MNRAS},
and the stellar mass and SFR of many high-$z$ galaxies  \citep{Navarro-Carrera2024, Wang2024ApJ}.
These superb data are combined with other high-$z$ observations from e.g. the Hubble Space Telescope (HST), the Spitzer telescope and the Atacama Large Millimeter/submillimeter Array (ALMA) telescope, which also measured the UVLF and stellar mass functions (SMF) of galaxies at $z > 6$ \citep{Bouwens2020, Bouwens2021, Stefanon2021}.

Both the 21-cm and high-$z$ galaxy observations will provide powerful probes for studying galaxy formation and reionization processes.
However, accurately modeling the complex physical processes during EoR remains challenging.
High-resolution cosmological hydrodynamic and radiative transfer simulations are able to model gas cooling, star formation, metal enrichment, feedback effects and gas ionizing processes, and thus consistently compute both galactic and IGM properties. {Among these simulations are SPHINX \citep{Rosdahl2018MNRAS, Katz2021MNRAS}, CoDa \citep{Ocvirk2016MNRAS, Ocvirk2020MNRAS}, CROC \citep{Gnedin2014ApJ, Gnedin2016ApJ}, THESAN \citep{Kannan2022, Borrow2023MNRAS, Kannan2025OJAp} and SPICE \citep{Bhagwat2024MNRAS}}, which are extremely computationally expensive and somewhat limited in terms of e.g. box size and/or mass resolution.
Thus, their computational costs inhibit a parameter space exploration.
In this respect, semi-analytical/numerical approaches such as MERAXES \citep{Mutch2016}, ASTRAEUS \citep{Hutter2021}, and POLAR \citep{Ma2023MNRAS} are more efficient in modeling the formation and evolution of galaxies and reionization, with the drawback that not all the physical processes are treated fully self-consistently.
21cmFAST \citep{Mesinger2011MNRAS} and {\sc GRIZZLY} \citep{Ghara2015MNRAS} apply semi-numerical calculations for the reionization process, but have no physical modeling of galaxy formation, and thus they can not naturally include e.g. the stochasticity of the UV luminosity versus halo mass relation \citep{Gelli2024ApJ, Nikolic2024A&A}.

Due to galaxy mergers and feedback effects, galaxies during the EoR may suffer diverse histories of starburst and quenching \citep{Furlanetto2022MNRAS, Sugimura2024ApJ}, and thus have very different star formation histories (SFH) \citep{Legrand2022MNRAS, Iyer2025arXiv}.
{Although the total ionizing photon emission of galaxies with the same stellar mass can be similar when evaluated over their entire history \citep[e.g.][]{Ma2025ApJ}, their different SFH can affect the recombination and cooling of ionized and heated IGM gas. Nevertheless, the effect of an evolving SFH  is typically not properly included in the semi-numerical approaches mentioned above to study the 21-cm signal during the EoR.
}
In this paper, we update POLAR \citep{Ma2023MNRAS, Acharya2025MNRAS} to explicitly include in the modeling of reionization the SFH of galaxies evaluated from semi-analytical galaxy formation simulations,
and explore how different SFHs affect the ionizing and heating processes and the 21-cm signal during the EoR.

The paper is organized as follows: we describe the galaxy formation, UV and X-ray sources, and radiative transfer models in Sect.~\ref{sec:method}, the results of high-$z$ galaxy SFH and reionization are presented in Sect.~\ref{sec:resul}, and the conclusions are summarized in Sect.~\ref{sec:conclu}.
The cosmological parameters adopted are $\Omega_{\Lambda} = 0.6889$, $\Omega_{m} = 0.3111$, $\Omega_{b} = 0.049$, $h = 0.6766$, $\sigma_{8} = 0.8102$ and $n_{s} = 0.9665$, from the {\it Planck} project \citep{Planck2020A&A} fitted with the data sets of TT, TE, and EE+lowE+lensing+BAO.

\section{Methods}
\label{sec:method}

\subsection{Galaxy formation and evolution}
Analogously to \cite{Ma2025ApJ}, to model galaxy formation and evolution, we use the {\sc Jiutian-300} {\it N}-body dark matter simulation \citep{Han2025SCPMA} in combination with the semi-analytic model {\sc L-Galaxies 2020} (named LG20 in the following; \citealt{Henriques_2020}).

{\sc Jiutian-300} was run with {\sc Gadget-4} \citep{Springel2021}, with a box size of $300\,h^{-1}\,{\rm cMpc}$ and $6144^{3}$ dark matter particles, i.e. particle mass $10^{7}\,h^{-1}\,{\rm M_{\odot}}$.
The simulation starts at $z=127$ and ends at $z=0$, with a total of 128 snapshot outputs, while we employ 39 of them at $z \ge 6$.
The dark matter halos are resolved with the Friend-of-Friend (FoF) algorithm \citep{Springel2001}, and the Subfind technique \citep{Springel2005} is applied to identify the sub-halos.
The halos have at least 20 dark-matter particles, i.e. the minimum halo mass is $2 \times 10^8\,h^{-1}\,{\rm M_{\odot}}$.
{Although to properly model high-$z$ galaxies and the large scale structure of the 21-cm signals during the EoR one should run simulations in hundreds of $\rm cMpc$ boxes at a much higher resolution than the one of {\sc Jiutan-300}, this is in practice not feasible from a computational point of view.
As a compromise, the mass resolution of {\sc Jiutian-300} is sufficient to evaluate the properties of high-$z$ galaxies \cite[see ][]{Ma2025ApJ}, and box size $300\,h^{-1}\,{\rm cMpc}$ is enough to compute the 21-cm signal on the relevant scales.}

LG20 is a semi-analytical galaxy formation model \citep{Henriques_2020}, which includes physical processes such as gas cooling, star formation, galaxy mergers, supernova and AGN feedback, chemical enrichment, tidal effects, and reincorporation of ejected gas.
It adopts a star formation law based on molecular hydrogen H$_{2}$.
We adopt the parameter values listed in the column LG20\_dust\_final of Table~1 in \cite{Ma2025ApJ}.
They are mostly those from \cite{Henriques_2020} consistent with observations of e.g. SMF at $z\le3$, while four parameters (i.e.  $\alpha_{\rm H_{2}}$, $\beta_{\rm SF,\,burst}$, $\gamma_{\rm reinc}$ and $M_{\rm r.p.}$) are the values that best fit UVLF observations of HST and JWST at $z=6-12$.
With this set of parameters, the LG20 results are consistent with UVLF and SMF measured at $z=6-12$.
We refer the reader to \cite{Ma2025ApJ} for more details about the parameter values and a comparison with observations.

Note that radiative feedback is not included in the galaxy formation model, as the radiative processes described in the following are treated in post-processing.
We note that star formation of very low-mass halos might be over-estimated without the inclusion of radiative feedback \citep{Hutter2021}.

\subsection{UV and X-ray sources}
Since the budget of ionizing photons is dominated by massive stars, while heating is dominated by X-ray binaries (XRBs) and shock heated ISM (hot-ISM) \citep{Eide2020MNRAS, Ma2021ApJ}, in the following we will discuss in more detail how the various sources are modeled.
Although the accretion of massive black holes (i.e. QSOs) can also emit abundant UV and X-ray photons, their total contribution to the EoR is usually considered to be sub-dominant due to their low number density \citep{Ma2021ApJ, Zeltyn2022ApJ, Jiang2022NatAs}, although some works suggest that QSOs might have a non-negligible contribution to reionization \citep{Ross2019MNRAS, Asthana2025MNRAS}.
In this work we do not include the impact of QSOs.

\subsubsection{Stellar sources}
For each galaxy at a specific $z$, LG20 saves the stellar and metal mass formed within $\sim 30$ time-bins along its history.
In each of them, we evaluate the Spectral Energy Distribution (SED) of the formed stellar mass with its age and metallicity using the binary star mode of Binary Population and Spectral Synthesis \cite[BPASS,][]{Eldridge2017, Stanway2018MNRAS}. We then integrate along the history to get the integrated SEDs (iSED).
As discussed in \cite{Ma2023MNRAS, Ma2025ApJ}, the iSED after normalization by stellar mass is not very sensitive to the redshift evolution, the stellar mass, and the galaxy formation model.

\begin{figure}
    \centering
    \includegraphics[width=0.95\linewidth]{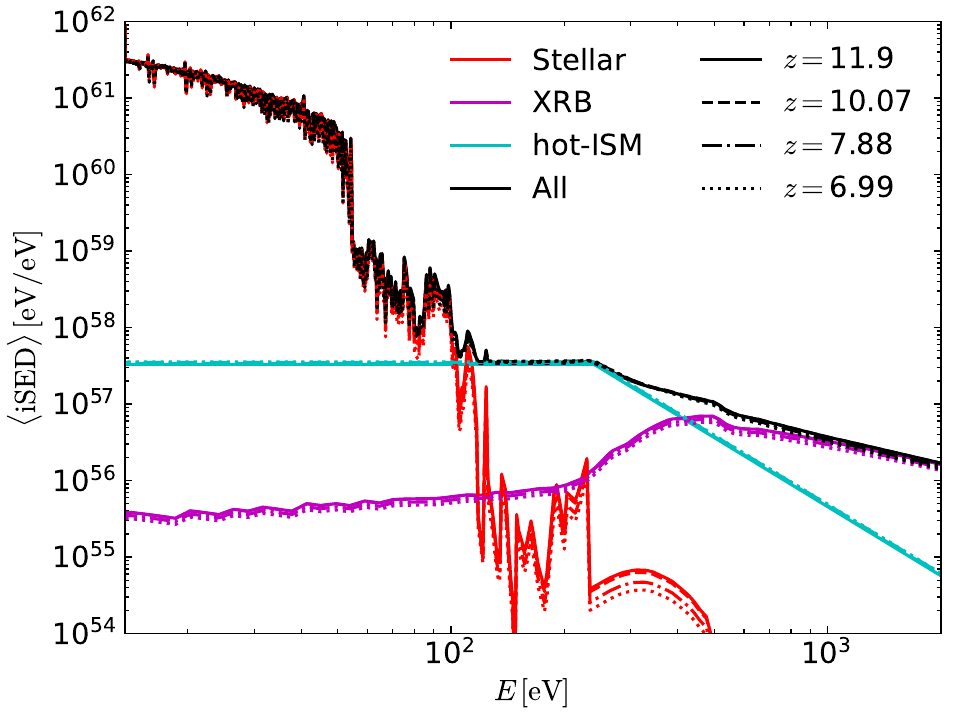}
    \caption {Average stellar mass normalized integrated SED, $\langle {\rm iSED} \rangle$, of stellar sources (red), XRB (magenta), hot-ISM (cyan) and all source types combined (black) at $z=11.9$ (solid), 10.07 (dashed), 7.88 (dash-dotted) and 6.99 (dotted) obtained from the LG20 simulation.
    These results are the mean values of all galaxies with $>10^{4}\,\rm M_{\odot}$.
    {The slight differences ($<10\%$) observed in $\langle {\rm iSED} \rangle$ at various redshifts are due to the evolution of galactic properties such as metallicity and stellar age.}
    }
    \label{fig:sed_full}
\end{figure}
As a reference, Fig.~\ref{fig:sed_full} shows the average iSED of stellar sources $\langle {\rm iSED} \rangle_{\rm stellar}$ after stellar mass normalization.
When compared to the total SED, obtained from the sum of all source types (denoted as ``All'' in Fig.~\ref{fig:sed_full}), we note that the stellar component dominates the UV radiation at $E<100\,\rm eV$, while it becomes negligible at higher energies.
Consistently with \cite{Ma2023MNRAS}, $\langle {\rm iSED} \rangle_{\rm stellar}$ has a negligible dependence on $z$, as we adopt a constant stellar IMF in the simulations and the stellar SED from the stellar population synthesis model \citep{Eldridge2017} is not sensitive to the metallicity evolution when the metallicity is below $Z_{\odot}$.

\subsubsection{X-ray binaries}
We model the luminosity and SED of X-ray binary systems (XRBs) employing the stellar mass and metallicity obtained from LG20 in combination with the scaling relations from \cite{Madau2017}.
For each galaxy, we compute the luminosity associated to its high-mass XRBs (HMXBs) $L_{\rm HMXB}$, and low-mass XRBs (LMXBs) $L_{\rm LMXB}$ within each time-bin along its SFH, where $L_{\rm HMXB}$ is related to the metallicity and SFR, and $L_{\rm LMXB}$ depends on the stellar age and mass.
The total XRB luminosity of one galaxy is then $L_{\rm XRB} = L_{\rm HMXB} + L_{\rm LMXB}$.
We adopt the SED template from \cite{Madau2017} to distribute the $L_{\rm XRB}$ into the energy range $13.6\,\rm eV - 2\,keV$.
Finally, we integrate along the SFH to obtain the integrated XRB SEDs for each galaxy.

In Fig.~\ref{fig:sed_full} we show the average stellar mass normalized iSED of XRBs $\langle {\rm iSED} \rangle_{\rm XRB}$ at four redshifts.
Similarly to $\langle {\rm iSED} \rangle_{\rm stellar}$, $\langle {\rm iSED} \rangle_{\rm XRB}$ also shows no redshift evolution.
We note that $\langle {\rm iSED} \rangle_{\rm XRB}$ dominates the X-ray emission at $E>400\,\rm eV$.

\subsubsection{Shock-heated hot-ISM}
The diffuse ISM shock-heated by supernova explosions (hot-ISM) can produce strong soft X-ray emission.
The luminosity of the hot-ISM is linearly proportional to the SFR with a factor $7.3 \pm 1.3 \times 10^{39}\,\rm erg\, s^{-1}\,M_{\odot}^{-1}\, yr$ within the energy range $0.3-10\,\rm keV$ \citep{Pacucci2014MNRAS}.
{Note that the linear relation of hot-ISM luminosity with SFR is due to the assumption of massive supernova explosion rate being proportional to the SFR, while such relation might be also dependent on the environment.
Since this is still not very clear, we just adopt the simplest linear relation previously adopted in other works.}
For each galaxy, we compute the luminosity emitted by the hot-ISM within each time-bin and then integrate along its SFH to obtain the integrated luminosity, which is then distributed within the energy range $13.6\,\rm eV - 2\,keV$ with the thermal bremsstrahlung spectral shape shown in \cite{Pacucci2014MNRAS}.

In Fig.~\ref{fig:sed_full}, we show the average stellar mass normalized iSED of hot-ISM $\langle {\rm iSED} \rangle_{\rm ISM}$ at four redshifts.
We see that $\langle {\rm iSED} \rangle_{\rm ISM}$ dominates the X-ray emission in the energy range $100 - 400\,\rm eV$, and that $\langle {\rm iSED} \rangle_{\rm ISM}$ shows no redshift evolution.

\subsection{Reionization and 21-cm signal}
\label{subsec:grizzly}
We use the one-dimensional (1-D) radiative transfer (RT) code {\sc Grizzly} \citep{Ghara2015MNRAS} to evaluate the ionization and temperature of the IGM, and the associated 21-cm signal.
Depending on the stellar mass and age of a galaxy, as well as on the underlying matter density,
{\sc Grizzly} first computes 1-D ionization and temperature profiles associated to each galaxy, and then the 3-D ionization and temperature maps by employing the FFT technique to include the ionizing and heating effects \citep{Ghara2015MNRAS}.
In the original version of {\sc Grizzly}, a constant SFR (defined as stellar mass divided by stellar age) is assumed.
Here, instead, we replace the SFH with constant SFR by the one obtained from LG20.
To make the computation more efficient, we take the average SFH of the galaxies within 29 stellar mass bins from $10^4$ to $10^{11}\,\rm M_{\odot}$ times 13 stellar age bins from $1$ to $10^{6}\,\rm Myr$.
We include the ionization and heating of UV and X-ray radiation from stellar sources, XRBs and hot-ISM by using  $\langle {\rm iSED} \rangle_{\rm All}$ shown in Fig.~\ref{fig:sed_full}.
We assume an escape fraction of 0.1 for UV photons ($E<100\,\rm eV$) and 1.0 for photons with $E>100\,\rm eV$.
During the latest stages of the EoR, when ionized regions merge into bigger ones, {\sc Grizzly} corrects for the effects of bubble overlap by conserving the ionizing photon budget.

{With the galaxy catalog from LG20 and the matter density maps from {\sc Jiutian-300}, we model the reionization of IGM with {\sc Grizzly}.
We run {\sc Grizzly} on a $256^3$ grid, i.e. the width of each cell is $1.17\,h^{-1}\,{\rm cMpc}$.
With such resolution, we expect the gas density to be roughly proportional to the dark matter density.
}
Finally, the 21-cm differential brightness temperature (DBT, $T_{\rm 21cm}$) is computed as
\begin{equation}
\label{eq:21cm}
\begin{split}
    T_{\rm 21cm} &= 27\,{\rm mK} \frac{\Omega_{\rm b}h^2}{0.023} \left(\frac{0.14}{\Omega_{\rm m}h^2}\frac{1+z}{10} \right)^{0.5} \\
    &\times (1+\delta_{\rm m})x_{\rm HI}\left(1-\frac{T_{\rm CMB}}{T_{\rm S}}\right)
\end{split}
\end{equation}
where $\delta_{\rm m}$ is the matter density obtained from the {\sc Jiutian-300} simulation, $x_{\rm HI}$ is the neutral hydrogen fraction computed by {\sc Grizzly}, $T_{\rm CMB}$ is the CMB temperature and $T_{\rm S}$ is the spin temperature.
We assume that in the redshift range of our interest, at $z<12$, a Ly$\alpha$ background has been established strong enough to couple the spin temperature to the kinetic temperature of the gas, i.e. $T_{\rm S} = T_{\rm k}$, where $T_{\rm k}$ is computed by {\sc Grizzly}.
Note that this assumption is not correct at very high-$z$.
To simplify the discussions, in the calculation of the 21-cm signal we also ignore redshift-space distortion (RSD) effects \citep{Mao2012MNRAS}.

\section{Results}
\label{sec:resul}
\subsection{Star formation history}
During its evolution, a galaxy might experience several starburst and quenching episodes due to physical processes such as gas cooling, galaxy merger, supernova and/or AGN feedback.
LG20 describes the SFH of a galaxy at $z$ in terms of its stellar mass weighted stellar age $\tau_{\rm age}(z) = \sum_{z'=z}^{z_{\rm start}} M_{\star}(z') t_{b}(z') /\sum_{z'=z}^{z_{\rm start}} M_{\star}(z')$, where $z_{\rm start}$ is the birth redshift of a galaxy, $M_{\star}(z')$ is the stellar mass formed within the time bin at $z'$ after the subtraction of mass loss due to supernova and AGB winds, and $t_{b}(z')$ is the time between $z$ and $z'$.
For galaxies with the same $M_{\star}$, a small $\tau_{\rm age}$ denotes more stars formed at a later time, while a larger $\tau_{\rm age}$ means that most stars in the galaxy are formed at an earlier time.

\begin{figure*}
    \centering
    \includegraphics[width=0.95\linewidth]{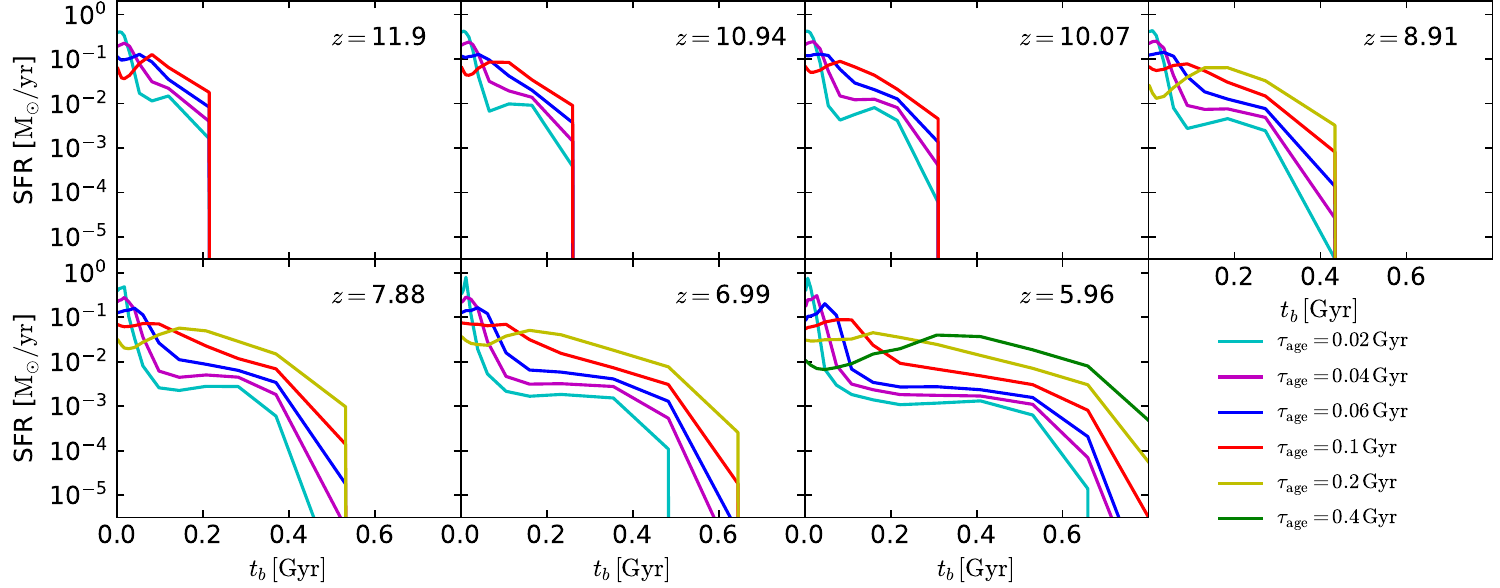}
    \caption{Average SFR history of galaxies with $M_{\star} \sim 10^{7}\,\rm M_{\odot}$ and stellar age $\tau_{\rm age}=0.02\,\rm Gyr$ (cyan), $0.04\,\rm Gyr$ (magenta), $0.06\,\rm Gyr$ (blue), $0.1\,\rm Gyr$ (red), $0.2\,\rm Gyr$ (yellow) and $0.4\,\rm Gyr$ (green) at $z=11.9$, 10.94, 10.07, 8.91, 7.88, 6.99 and 5.96 from left to right and top to bottom.
    The x-axis $t_{b}$ is the time of galaxies traced back from $z$ to higher redshift.
    }
    \label{fig:sfh_mulitz}
\end{figure*}
To present how the SFH of galaxies is sensitive to $\tau_{\rm age}$, Fig.~\ref{fig:sfh_mulitz} shows the history of average SFR of galaxies with $M_{\star} \sim 10^{7}\,\rm M_{\odot}$ but with different $\tau_{\rm age}$ at seven redshifts from 12 to 6.
{The SFH with 1-$\sigma$ scatter of Fig.~\ref{fig:sfh_mulitz} is shown in Appendix~\ref{SFH_scatter}.
Despite the fairly large scatter, we can still observe some obvious differences for galaxies with different $\tau_{\rm age}$.
}
Note that, since at very high-$z$ no galaxy is very massive and has a large $\tau_{\rm age}$, these lines are not shown in the plots.
We observe that the SFH of galaxies is highly dependent on $\tau_{\rm age}$, and it also evolves with $z$, although there are some common features with the same $\tau_{\rm age}$.
More specifically, the SFR of galaxies roughly peaks at $t_{b} \approx \tau_{\rm age}$, e.g. galaxies with $\tau_{\rm age} = 0.4\,\rm Gyr$ at $z=5.96$ have the highest SFR at $t_{b} \sim 0.4 \,\rm Gyr$.
This conclusion is redshift independent, although galaxies at high-$z$ (e.g. $z=11.9$) have much shorter star formation period than those at low $z$ (e.g. $z=5.96$).
As $\tau_{\rm age}$ decreases, the SFR becomes higher at $t_{b} \sim 0 \,\rm Gyr$, e.g. galaxies with $\tau_{\rm age} = 0.02\,\rm Gyr$ show highest SFR at $t_{b} \sim 0 \,\rm Gyr$.

\begin{figure}
    \centering
    \includegraphics[width=0.95\linewidth]{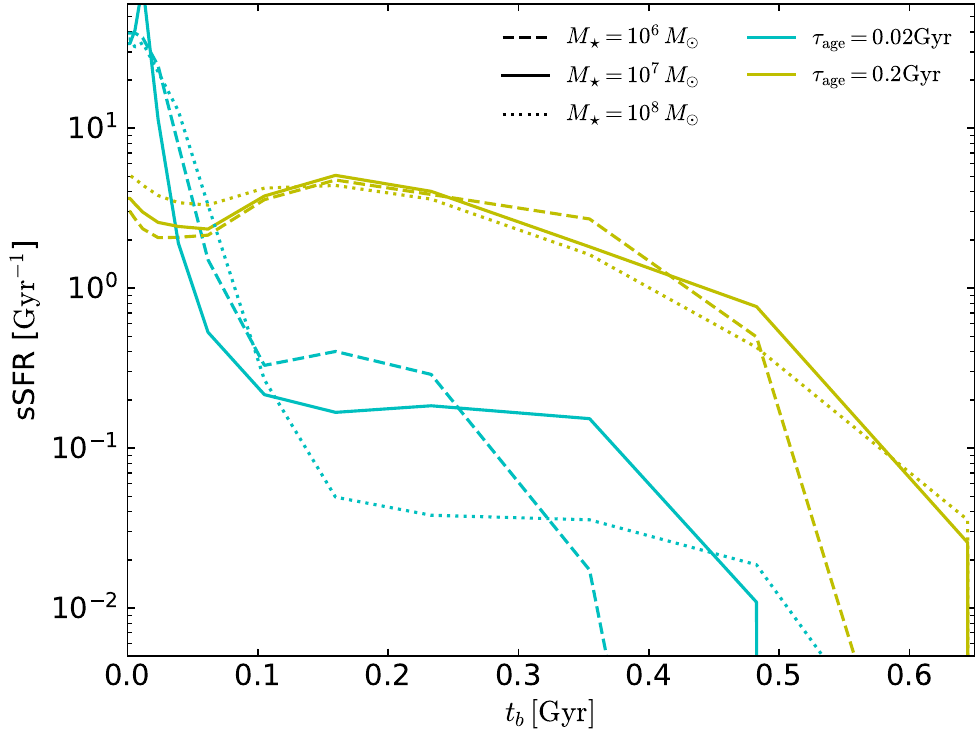}
    \caption{Average specific SFR (i.e. SFR per unit stellar mass) history of galaxies with {stellar age $\tau_{\rm age}=0.02\,\rm Gyr$ (cyan) and $0.2\,\rm Gyr$ (yellow)} for galaxies with $M_{\star} \sim 10^{6}\,\rm M_{\odot}$ (dashed), $10^{7}\,\rm M_{\odot}$ (solid) and $10^{8}\,\rm M_{\odot}$ (dotted).
    The results are at $z=6.99$.
    }
    \label{fig:sfh_z7_mass}
\end{figure}
Figure~\ref{fig:sfh_z7_mass} shows the histories of the specific SFR (sSFR, i.e. the SFR divided by the stellar mass) of galaxies with different $M_{\star}$ and $\tau_{\rm age}$ at $z=6.99$.
Consistently with Fig.~\ref{fig:sfh_mulitz}, the sSFR of galaxies is sensitive to $\tau_{\rm age}$, while it does not depend very much on the stellar mass $M_{\star}$.
More specifically, galaxies with the same $\tau_{\rm age}$ but different $M_{\star}$ have roughly similar sSFR histories, e.g. the sSFR of galaxies with three $M_{\star}$ are similar at $t_{b} < 0.5\,\rm Gyr$ in the case of $\tau_{\rm age} = 0.2\,\rm Gyr$.
Differently, with a fixed $\tau_{\rm age}$, more massive galaxies can have earlier star formation than the less massive ones.

\begin{figure*}
    \centering
    \includegraphics[width=0.95\linewidth]{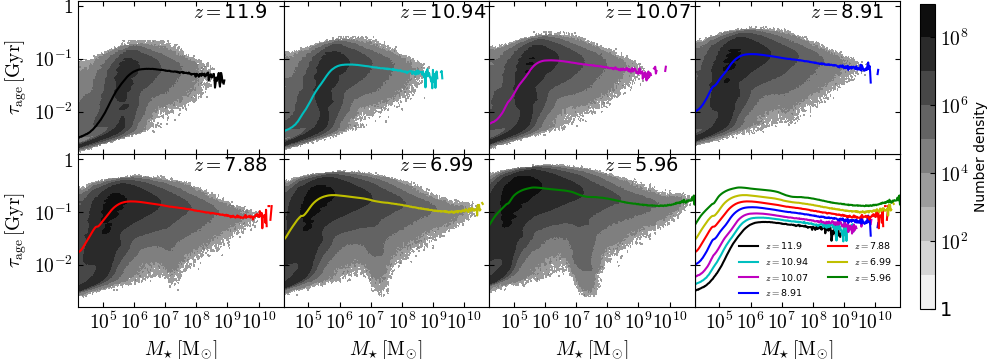}
    \caption{Distributions of stellar age $\tau_{\rm age}$ versus stellar mass $M_{\star}$ of galaxies at $z=11.9$, 10.94, 10.07, 8.91, 7.88, 6.99 and 5.96, from left to right and top to bottom.
    The solid lines are the mean $\tau_{\rm age}$ of galaxies within the same $M_{\star}$ bins, which are shown together for all redshifts in the bottom-right plot.}
    \label{fig:age_vs_mass}
\end{figure*}
Figure~\ref{fig:age_vs_mass} shows the distributions of stellar mass weighted stellar age $\tau_{\rm age}$ versus stellar mass $M_{\star}$ of galaxies at various redshifts.
At the same $z$, the mean $\tau_{\rm age}$ increases with increasing $M_{\star}$ for $M_{\star}<10^{6}\,\rm M_{\odot}$, while it decreases for $M_{\star}>10^{6}\,\rm M_{\odot}$.
This is due to the higher SFR in the massive galaxies, while star formation in smaller objects ($\sim 10^{6}\,\rm M_{\odot}$) is easily quenched by supernova feedback.
With the same $M_{\star}$, the mean $\tau_{\rm age}$ of galaxies increases with decreasing $z$.
We note that the number density of galaxies peaks at $10^{5}\,{\rm M_{\odot}}<M_{\star}<10^{6}\,\rm M_{\odot}$, while the lack of lower mass galaxies is due to a combination of the resolution limit of {\sc Jiutian-300} and the supernova feedback effect.
As shown in \cite{Ma2025ApJ}, the galaxy sample at $M_{\star}<10^{7}\,\rm M_{\odot}$ might be incomplete due to the mass resolution of the simulation, and thus the relation of $\tau_{\rm age}$ and $M_{\star}$ might not be very robust at $M_{\star}<10^{7}\,\rm M_{\odot}$.
As the $\tau_{\rm age}$ and $M_{\star}$ relations at all redshifts have very strong scatters, in the {\sc Grizzly} simulations we take $\tau_{\rm age}$ and $M_{\star}$ as two independent properties for each galaxy.

\subsection{Reionization}
Using the galactic stellar mass $M_{\star}$, stellar age $\tau_{\rm age}$ and SFH obtained from LG20 as input quantities, we employ {\sc Grizzly} to compute the reionization process and the associated 21-cm signal.
To investigate how the ionization and heating of the IGM and the associated 21-cm signal from the EoR depend on the galactic SFH, we run six {\sc Grizzly} simulations with different SFH models.
As a reference, three of them adopt a constant stellar age, i.e. $\tau_{\rm age} = 0.02\,\rm Gyr$, $0.1\,\rm Gyr$ and $0.2\,\rm Gyr$, while one uses the $\tau_{\rm age}$ of galaxies obtained from LG20. The simulations are named simul\_0.02, simul\_0.1, simul\_0.2 and simul\_Fiducial, respectively.
In all simulations, the galactic SFH is the one obtained from LG20.
Additionally, for simul\_0.1 and simul\_Fiducial we also run a case in which the SFR is kept constant (i.e. SFR $= M_{\star} / \tau_{\rm age}$) along the galactic history with period of $\tau_{\rm age}$. These simulations are named simul\_0.1\_const and simul\_Fiducial\_const, respectively.
In the following we will employ these six {\sc Grizzly} simulations to study the effects of different modeling of SFH on the ionizing and heating processes and on the 21-cm signal during EoR.

\subsubsection{Ionization and gas temperature}
\begin{figure*}
    \centering
    \includegraphics[width=0.95\linewidth]{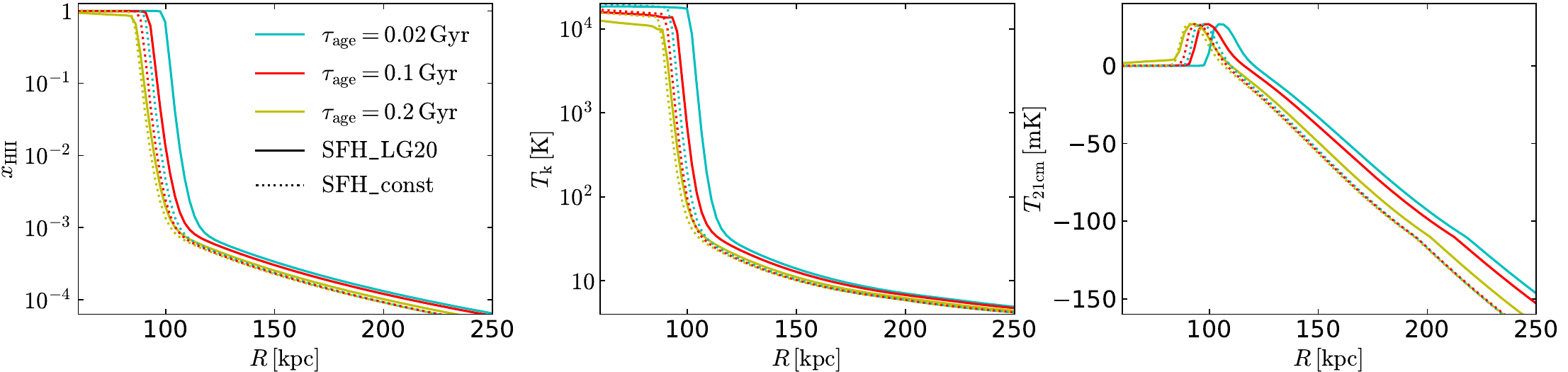}
    \caption{1-D profiles of ionization fraction $x_{\rm HII}$ (left), gas temperature $T_{\rm k}$ (middle) and 21-cm signal $T_{\rm 21cm}$ (right) as functions of the physical distance $R$ from a galaxy with $M_{\ast} = 10^7\,\rm M_{\odot}$ at $z=10.07$.
    The colors refer to galaxies with $\tau_{\rm age} = 0.02\,\rm Gyr$ (cyan), $0.1\,\rm Gyr$ (red) and  $0.2\,\rm Gyr$ (yellow).
    The solid lines are the results obtained by adopting the SFH from LG20  ($\rm SFH\_LG20$), while the dotted lines refer to a constant SFR throughout the SFH ($\rm SFH\_const$).
    The surrounding IGM is assumed to be uniform with mean density of the Universe at $z=10.07$.
    }
    \label{fig:s21cm_1d_theo}
\end{figure*}
As mentioned in Sect.~\ref{subsec:grizzly}, {\sc Grizzly} first pre-computes the 1-D ionization and temperature profiles along the radius away from galaxies with different stellar mass, stellar age and IGM matter density.
As a reference, Fig.~\ref{fig:s21cm_1d_theo} shows some samples of ionization fraction $x_{\rm HII}$, gas temperature $T_{\rm k}$ and 21-cm DBT $T_{\rm 21cm}$ profiles for galaxies with the same stellar mass $M_{\ast}$ but different stellar age $\tau_{\rm age}$ and SFH models at $z=10.07$.
Galaxies with $M_{\ast}=10^7\,\rm M_{\odot}$ but different $\tau_{\rm age}$ have similar ionized bubbles, of extension $\sim 90\,\rm kpc$, as the ionized photon budget is roughly linearly proportional to $M_{\ast}$ \citep{Ma2025ApJ}. Some differences, though, are visible.
More specifically, the radius of the ionized bubble becomes slightly smaller with increasing $\tau_{\rm age}$. For example, for $\tau_{\rm age} = 0.2\,\rm Gyr$ the radius is $\sim 13\%$ smaller (corresponding to $\sim 34\%$ smaller in ionized volume) than for $\tau_{\rm age} = 0.02\,\rm Gyr$, because the recombination of ionized hydrogen and free electrons becomes more significant for longer $\tau_{\rm age}$.
Meanwhile, for the same $\tau_{\rm age} = 0.02\,\rm Gyr$, the ionized bubble of a galaxy with SFH from LG20 is roughly $\sim 10\%$ larger than the one with a constant SFR. The difference reduces to $\sim 5\%$  and $\sim 1\%$ for $\tau_{\rm age} = 0.1\,\rm Gyr$ and $0.2\,\rm Gyr$, respectively.
The tail of low ionization visible at larger distances beyond the fully ionized bubble is due to the high energy photons which can only partially ionize the neutral gas \cite[see also][]{Ghara2015MNRAS}, reaching e.g. $x_{\rm HII} = 10^{-4}$ at $R \sim 200\, \rm kpc$.
The profiles of the gas temperature $T_{\rm k}$ are similar to those of $x_{\rm HII}$, i.e.  $T_{\rm k}$ becomes slightly smaller as $\tau_{\rm age}$ increases, while for the same $\tau_{\rm age}$ the galaxy with SFH from LG20 has higher $T_{\rm k}$ than the one with constant SFR.
The differences of $x_{\rm HII}$ and $T_{\rm k}$ caused by the various modeling of SFH also result in obvious differences in the 21-cm DBT $T_{\rm 21cm}$.
Since $T_{\rm 21cm} \propto (1-T_{\rm CMB}/T_{\rm S})$ (see Eq.~\ref{eq:21cm}), such differences are more significant at $R>120\, \rm kpc$, where the gas temperature $T_{\rm k}$ is below $T_{\rm CMB}$.

\begin{figure*}
    \centering
    \includegraphics[width=0.95\linewidth]{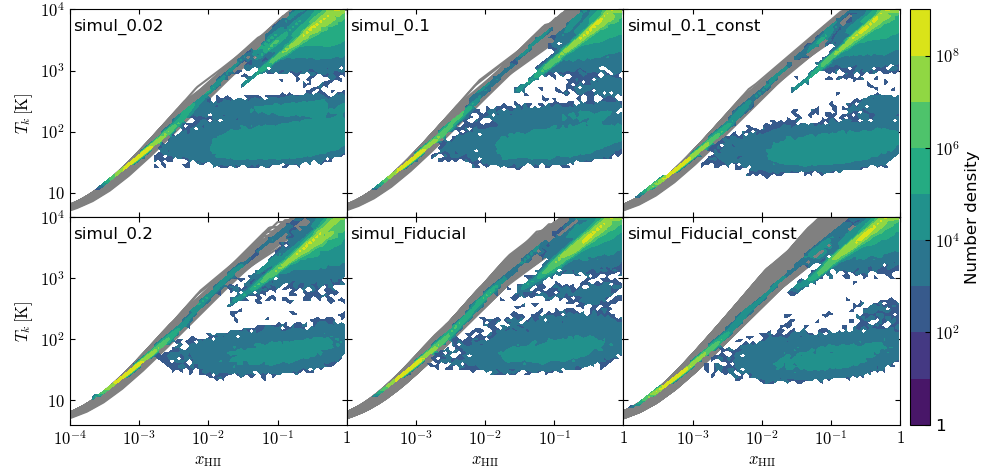}
    \caption{2-D distributions of gas temperature $T_{k}$ versus ionization fraction $x_{\rm HII}$ from simulation simul\_0.02, simul\_0.1, simul\_0.1\_const, simul\_0.2, simul\_Fiducial and simul\_Fiducial\_const at $z=10.07$, from left to right and top to bottom.
    simul\_0.1\_const and simul\_Fiducial\_const have a constant SFR, while the others adopt the SFH obtained from LG20.
    The background gray lines are the $T_{k}$ as functions of $x_{\rm HII}$ from the 1-D RT calculations.
    }
    \label{fig:temp_vs_xhii_simul}
\end{figure*}
To investigate how the different modeling of SFH affects the ionization and gas temperature of the IGM, in Fig.~\ref{fig:temp_vs_xhii_simul} we show the 2-D distributions of gas temperature $T_{k}$ as a function of the ionization fraction $x_{\rm HII}$ at $z=10.07$.
As a reference, we also show the $T_{k}$ as functions of $x_{\rm HII}$ from the 1-D RT calculations for galaxies with different stellar mass, stellar age and matter density which are pre-computed by {\sc Grizzly}.
Note that the size of the ionized bubbles can be larger than the one estimated by the 1-D RT calculations due to the correction applied to take into account the overlap of ionized regions \cite[see][for more details]{Ghara2015MNRAS}.
This means that in the simulation box many cells with the same $T_{k}$ show higher $x_{\rm HII}$ than the ones from the 1-D RT calculations.
In all simulations, the IGM cells are roughly located in three regions.
The first is at $T_{k}>10^{3}\,\rm K$ with $x_{\rm HII} > 0.05$, comprising gas heated and ionized by both UV and X-ray photons.
Since the $T_{k}$ of these cells is much higher than the CMB temperature, the slight differences shown in the six simulations would not be visible in the corresponding 21-cm signal.
The second region is at $T_{k} < 400\, K$ with $x_{\rm HII} > 0.003$, where the cells have $x_{\rm HII}$ higher than those from the 1-D RT calculations due to the correction for overlap.
With constant $\tau_{\rm age}$ and SFH from LG20, the $T_{k}$ of cells within this region becomes lower with increasing $\tau_{\rm age}$. For example, their $T_{k}$ from simul\_0.2 is typically lower than the one from simul\_0.02, consistently with  Fig.~\ref{fig:s21cm_1d_theo}.
The $T_{k}$ within this region from simulations with SFH from LG20 is higher than the one with a constant SFR. For example, the one from simul\_0.1 and simul\_Fiducial is higher than from simul\_0.1\_const and simul\_Fiducial\_const, respectively.
Since most cells within this region have $T_{k} < 100\,$K, we expect to see differences in the associated 21-cm signal of simulations with different SFH models.
In the third region the $T_{k}$ and $x_{\rm HII}$ values overlap with those from the 1-D RT calculation. These are cells located far away from the ionization-fronts (I-front) and thus are not affected by the correction for overlap.
Many cells within this region also have $T_{k} < 100\,$K, and the $T_{k}$ values decrease with increasing $\tau_{\rm age}$ in the simulations with constant $\tau_{\rm age}$ and SFH from LG20, and the ones from simulations with SFH from LG20 are higher than those with constant SFR.
Such differences of $T_{k}$ within this region from simulations with different SFH models are expected to affect the corresponding 21-cm signals.

\subsubsection{The 21-cm signal}
\begin{figure*}
    \centering
    \includegraphics[width=0.95\linewidth]{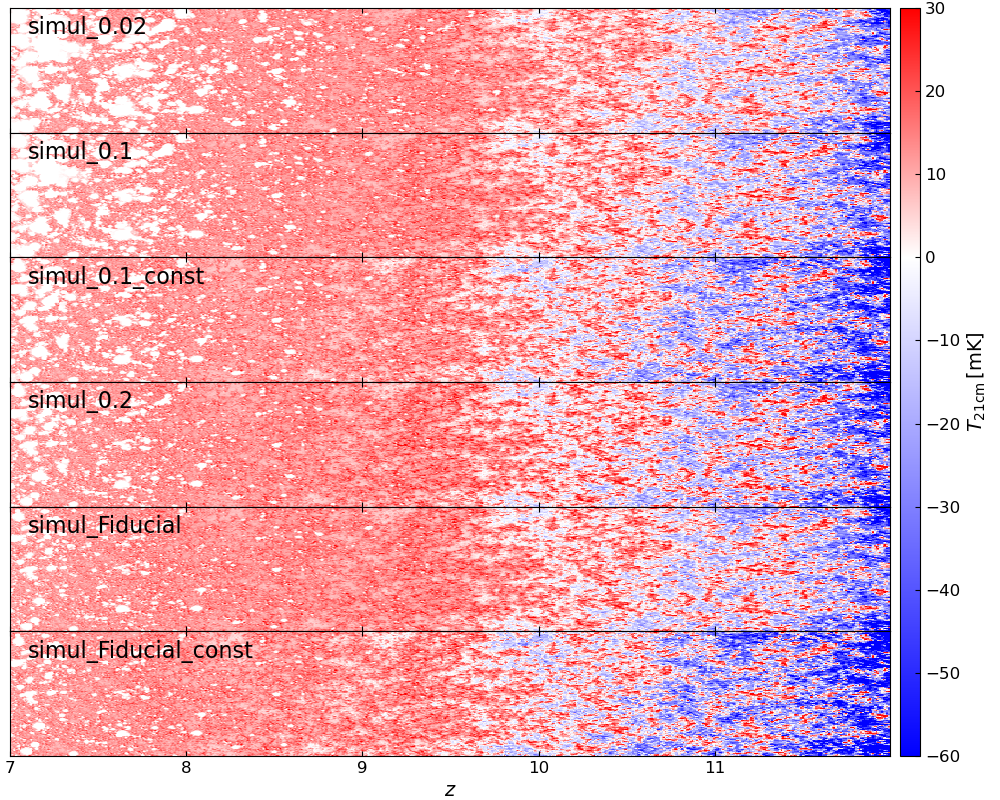}
    \caption{Light-cones of 21-cm DBT $T_{\rm 21cm}$ from $z=7$ to 12 extracted from simulation simul\_0.02, simul\_0.1, simul\_0.1\_const, simul\_0.2, simul\_Fiducial and simul\_Fiducial\_const from top to bottom.
    }
    \label{fig:image_21cm}
\end{figure*}
Figure~\ref{fig:image_21cm} shows the light-cones of $T_{\rm 21cm}$ extracted from six simulations.
At $z=12$, most of the 21-cm signal is still in absorption (blue cells), and fully ionized gas (white cells) is negligible.
Because of the heating from the X-ray sources (see Fig.~\ref{fig:sed_full}), the 21-cm signal becomes mostly in emission at $z \sim 10$.
Ionized bubbles are clearly visible at $z=7$, where all the cells are highly heated and the 21-cm signal is in emission.
The simulations with different modeling of SFH show some visible differences.
simul\_0.02 has more ionization at $z=7$ and higher $T_{\rm 21cm}$ at $z=12$ than simul\_0.2, due to the larger ionized bubbles and stronger gas heating produced by the galaxies with smaller $\tau_{\rm age}$ (see Fig.~\ref{fig:s21cm_1d_theo}).
As galaxies with SFH from LG20 can produce larger ionized bubbles than those with constant SFR (see Fig.~\ref{fig:s21cm_1d_theo}), simul\_0.1 has more ionization at $z=7$ than simul\_0.1\_const.
simul\_Fiducial shows an ionization history similar to the one from simul\_Fiducial\_const, while the former has higher $T_{\rm 21cm}$ than the latter at e.g. $z>10$.

\begin{figure*}
    \centering
    \includegraphics[width=0.48\linewidth]{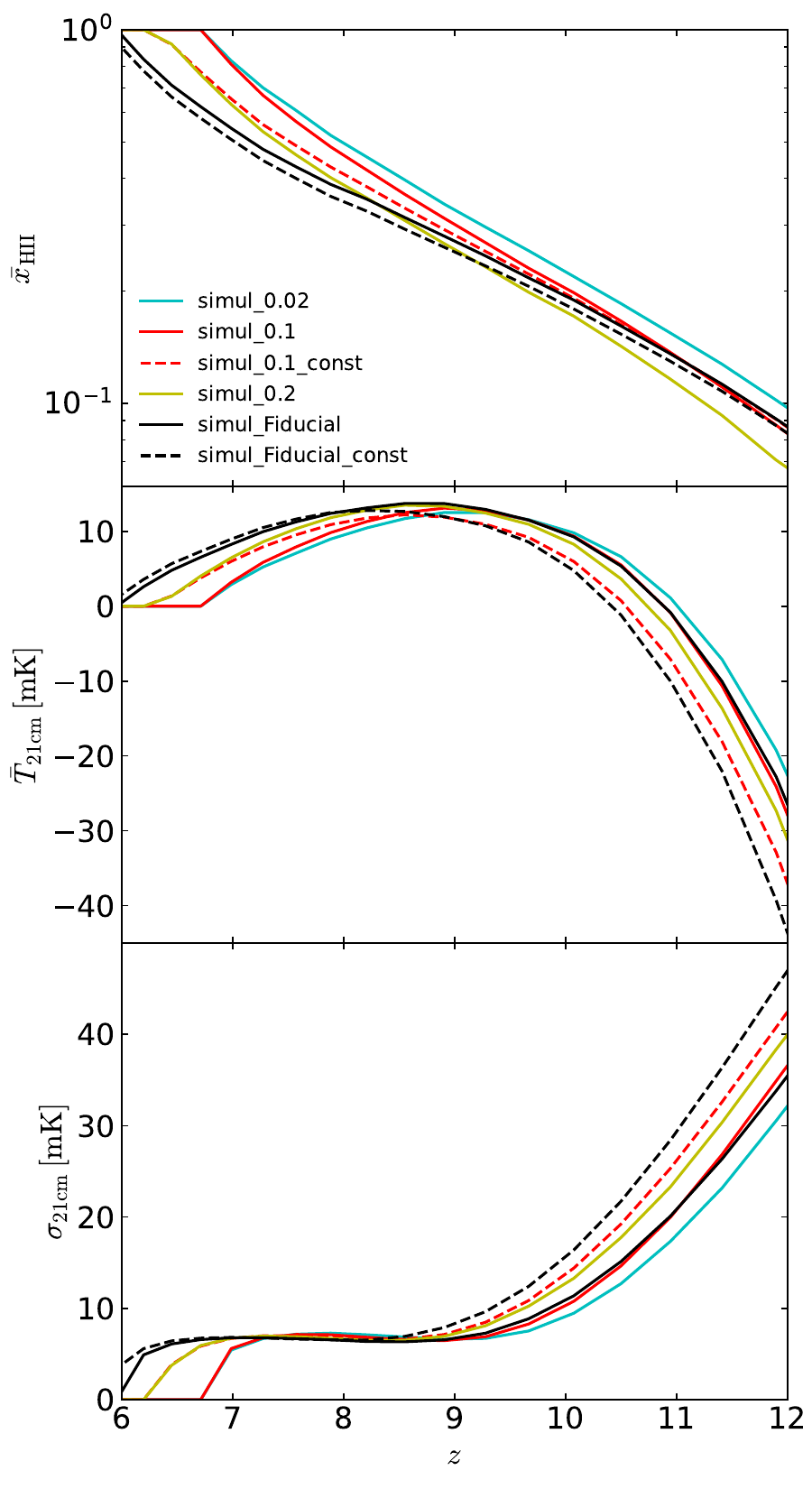}
    \includegraphics[width=0.48\linewidth]{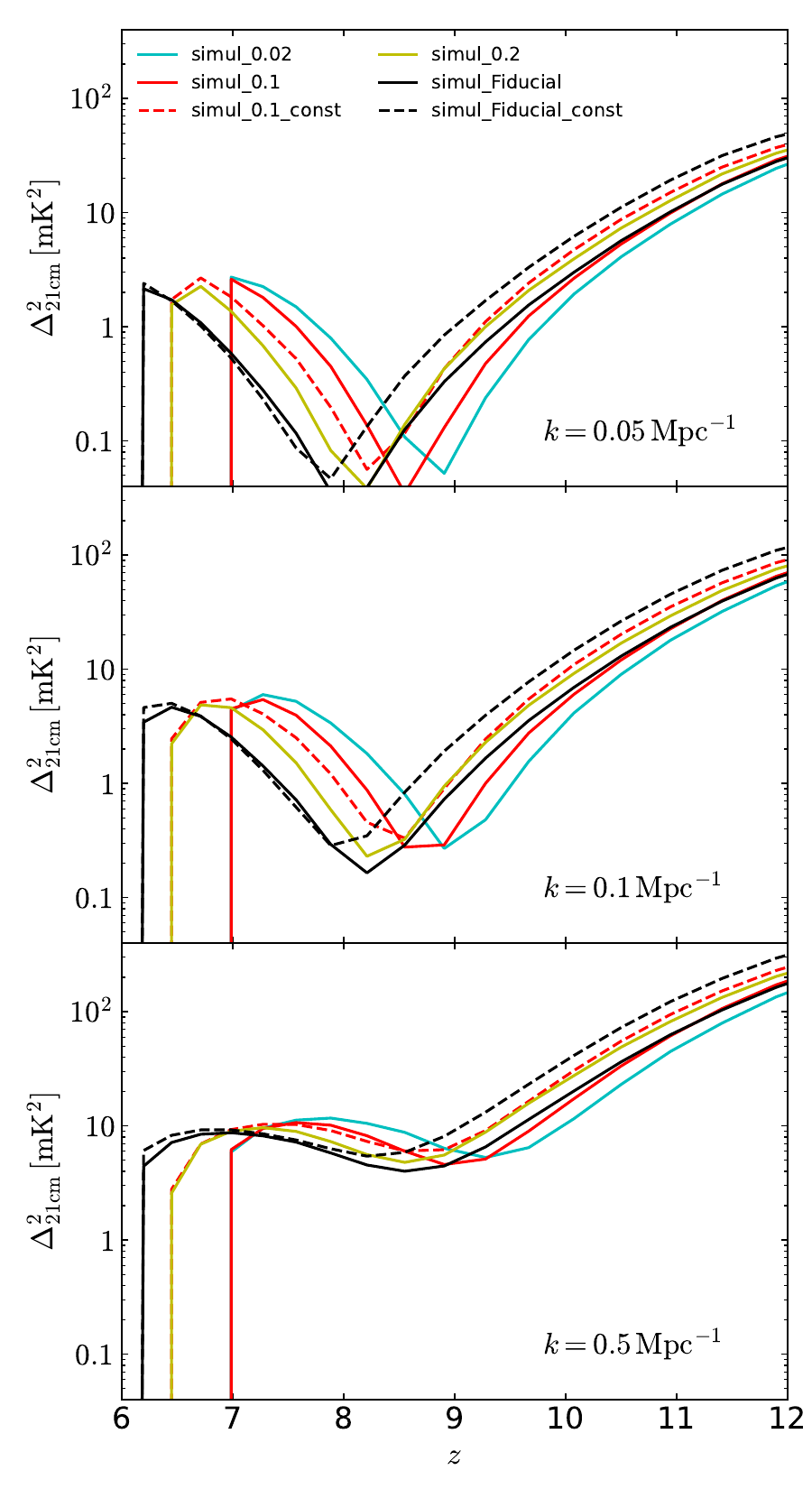}
    \caption{{\bf Left}: histories of volume-averaged mean ionization fraction $\bar{x}_{\rm HII}$ (top), mean 21-cm DBT $\bar{T}_{\rm 21cm}$ (central) and rms of 21-cm DBT $\sigma_{\rm 21cm}$ (bottom) from simulation simul\_0.02 (solid cyan), simul\_0.1 (solid red), simul\_0.1\_const (dashed red), simul\_0.2 (solid yellow), simul\_Fiducial (solid black) and simul\_Fiducial\_const (dashed black).
    {\bf Right}: redshift evolution of 21-cm power spectra $\Delta^2_{\rm 21cm}$ at $k=0.05\,\rm Mpc^{-1}$ (top), $0.1\,\rm Mpc^{-1}$ (central), and $0.5\,\rm Mpc^{-1}$ (bottom) from the same simulations.}
    \label{fig:xhii_vz_evol}
\end{figure*}
The left panel of Fig.~\ref{fig:xhii_vz_evol} shows the evolution of the volume-averaged mean ionization fraction $\bar{x}_{\rm HII}$, mean 21-cm DBT $\bar{T}_{\rm 21cm}$ and rms (root mean square) of 21-cm DBT $\sigma_{\rm 21cm}$ for different modeling of SFH.
As shown in Fig.~\ref{fig:s21cm_1d_theo}, galaxies with smaller $\tau_{\rm age}$ produce larger ionized bubbles, thus the EoR e.g. in simul\_0.02 ends $\Delta z \approx 0.5$ earlier than in simul\_0.2.
The EoR in simul\_0.1 finishes earlier than in simul\_0.1\_const, as the galaxies with a SFH from LG20 produce ionized bubbles larger than those with a constant SFR, as shown in Fig.~\ref{fig:s21cm_1d_theo}.
For the same reason, in simul\_Fiducial the EoR ends slightly earlier than in simul\_Fiducial\_const.
At $z>10$, simul\_Fiducial, simul\_Fiducial\_const, simul\_0.1 and simul\_0.1\_const have similar $\bar{x}_{\rm HII}$, while simul\_0.02 and simul\_0.2 produce a higher and lower $\bar{x}_{\rm HII}$ at the same $z$, respectively.

Galaxies with the same $M_{\star}$ but different modeling of SFH can exhibit different ionization and heating patterns (see Fig.~\ref{fig:s21cm_1d_theo}), and thus associated 21-cm signatures, in particular at $z>9$ when heating dominates the signal (see also the gas temperature distributions shown in Fig.~\ref{fig:temp_vs_xhii_simul}).
More specifically, a higher $\tau_{\rm age}$ means less X-ray heating, and thus simul\_0.2 has a $\bar{T}_{\rm 21cm}$ lower than simul\_0.1 and simul\_0.02.
The $\bar{T}_{\rm 21cm}$ of simul\_0.1, instead, is higher than that of simul\_0.1\_const, because of the larger heating from galaxies with SFH obtained from LG20 in comparison to those with a constant SFR (see also Fig.~\ref{fig:temp_vs_xhii_simul}).
For the same reason, simul\_Fiducial has $\bar{T}_{\rm 21cm}$ higher than  simul\_Fiducial\_const.
$\sigma_{\rm 21cm}$ is roughly related to the absolute value of $\bar{T}_{\rm 21cm}$, and a higher absolute value of $\bar{T}_{\rm 21cm}$ leads to a larger $\sigma_{\rm 21cm}$.
When the neutral gas becomes highly heated (e.g. $z<9$), $\bar{T}_{\rm 21cm}$ is dominated by the contribution from the ionization fraction $\bar{x}_{\rm HII}$, and a higher $\bar{x}_{\rm HII}$ leads to a lower $\bar{T}_{\rm 21cm}$ and $\sigma_{\rm 21cm}$.

The right panel of Fig.~\ref{fig:xhii_vz_evol} shows the evolution of the 21-cm power spectra $\Delta^2_{\rm 21cm}$ at $k=0.05\,\rm Mpc^{-1}$, $0.1\,\rm Mpc^{-1}$, and $0.5\,\rm Mpc^{-1}$ for simulations with different modeling of the SFH.
All power spectra exhibit similar features at the three $k$-values, i.e. they decrease with decreasing $z$ at $z>9$ due to the X-ray heating, while at $z<8$ they increase with decreasing $z$ due to the inhomogeneous ionizing process.
The modeling of SFH also imprints some differences on $\Delta^2_{\rm 21cm}$.
At $z>9$, when the bulk of the IGM gas is only weakly heated, X-ray heating can increase the local 21-cm fluctuations. For this reason, $\Delta^2_{\rm 21cm}$ is higher in simulations with less gas heating, and its evolution roughly follows the $\sigma_{\rm 21cm}$ shown in the left panel of Fig.~\ref{fig:xhii_vz_evol}.
More specifically, simul\_0.2 has $\Delta^2_{\rm 21cm}$ higher than simul\_0.1 and simul\_0.02, as the weaker X-ray heating of galaxies with $\tau_{\rm age} = 0.2\,\rm Gyr$ (see Fig.~\ref{fig:s21cm_1d_theo}) leads to smaller $T_{\rm 21cm}$ but higher $\sigma_{\rm 21cm}$ and $\Delta^2_{\rm 21cm}$.
Since the heating of galaxies with SFH from LG20 is more than from those with constant SFR, the $\Delta^2_{\rm 21cm}$ of simul\_0.1 and simul\_Fiducial is lower than simul\_0.1\_const and simul\_Fiducial\_const, respectively.
At $z<8$ peaks in the $\Delta^2_{\rm 21cm}$ emerge before the end of the EoR, when the neutral gas is highly heated and the 21-cm fluctuations are dominated by the ionized bubbles.
For example, the $\Delta^2_{\rm 21cm}$ of simul\_0.2 peaks at a $z$ ($\sim6.6$) lower than those of simul\_0.1 and simul\_0.02, while with the SFH from LG20 the $\Delta^2_{\rm 21cm}$ of  simul\_0.1 peaks at a $z$ higher than that of simul\_0.1\_const with constant SFR.
simul\_Fiducial and simul\_Fiducial\_const have similar $\Delta^2_{\rm 21cm}$ at $z<8$, consistently with their $\bar{x}_{\rm HII}$ and $\bar{T}_{\rm 21cm}$ shown in the left panel of Fig.~\ref{fig:xhii_vz_evol}.

\section{Conclusions and Discussions}
\label{sec:conclu}
A wealth of data from the Epoch of Reionization (EoR) has been recently (or will soon be) obtained, e.g. on galaxy properties by JWST, and on the 21-cm signal from the IGM gas by LOFAR, NenuFAR, HERA, MWA and SKA.
To optimally exploit these data, we have developed the semi-numerical code POLAR \citep{Ma2023MNRAS, Acharya2025MNRAS}, which consistently and efficiently models the galaxy formation and reionization processes.
In this work, we update POLAR to include the effect of an evolving star formation history (SFH) in the modeling of reionization, as well as of source populations other than stars.
We investigate how these affect IGM heating and ionization, as well as the 21-cm signal during the EoR.

We adopt the {\sc Jiutian-300} {\it N}-body dark matter simulation \citep{Han2025SCPMA}, which can resolve a minimum halo mass of $2 \times 10^8\,h^{-1}\,{\rm M_{\odot}}$, and the semi-analytic model {\sc L-Galaxies 2020} (LG20; \citealt{Henriques_2020}), to follow the formation and evolution of galaxies.
We then post-process the star formation and metallicity histories of each galaxy to obtain the luminosity and spectral energy distributions (SED) of stellar sources, X-ray binaries (XRB) and shock-heated ISM (hot-ISM).
Finally, we employ the 1-D radiative transfer (RT) code {\sc Grizzly} together with the galaxy catalog from LG20 and the matter density maps from {\sc Jiutian-300} to model the IGM reionization history and the associated 21-cm signal.

We find that the stellar mass weighted stellar age $\tau_{\rm age}$ of galaxies with the same $M_\star$ is very sensitive to the SFH, and that galaxies with the same $\tau_{\rm age}$ but different stellar mass have similar specific SFR.
The mean values of $\tau_{\rm age}$ decrease with increasing $M_{\star}$ at $M_{\star}>10^{6}\,\rm M_{\odot}$, while for a fixed $M_{\star}$ the mean $\tau_{\rm age}$ increases with decreasing $z$.
We also find that for galaxies with the same stellar mass, a large $\tau_{\rm age}$ results in more recombination of ionized hydrogen and free electrons, and thus smaller ionized regions than those produced by galaxies with smaller $\tau_{\rm age}$.
For the same $\tau_{\rm age}$ and stellar mass, the ionized bubbles surrounding galaxies with SFH obtained from LG20 are slightly larger than those with a constant SFR, although differences become smaller with increasing $\tau_{\rm age}$.
The different modeling of SFH also leads to slightly different partial ionization and gas temperature profiles due to the X-ray heating, which can consequently result in  differences in the 21-cm differential brightness temperature (DBT) $T_{\rm 21cm}$.

We run six RT {\sc Grizzly} simulations with different modeling of the SFH to explore their impact on the 21-cm signal during the EoR.
{All the simulations have the same ionizing photon budget, as the total ionizing photon emission of galaxies with the same stellar mass is similar over their entire history \citep{Ma2025ApJ}.}
We find that varying the modeling of SFH can change the dependence of the gas temperature on the ionization fraction.
With the SFH obtained from LG20 the simulations with larger constant $\tau_{\rm age}$ lead to lower gas temperatures, and the gas temperature of simulations with SFH from LG20 is higher than those with constant SFR.
Since galaxies with smaller $\tau_{\rm age}$ have larger ionized bubbles, the simulation with small constant $\tau_{\rm age}$ ends the reionization process earlier than the one with larger constant $\tau_{\rm age}$.
For the same $\tau_{\rm age}$, the simulations with a SFH from LG20 finish the EoR earlier than those with constant SFR.
These features also appear in the evolution of the 21-cm power spectra $\Delta^2_{\rm 21cm}$, e.g. at $z>9$ the simulation with a large constant $\tau_{\rm age}$ has $\Delta^2_{\rm 21cm}$ higher than the one with smaller constant $\tau_{\rm age}$, and the simulations with SFH from LG20 have $\Delta^2_{\rm 21cm}$ lower than the ones with constant SFR.

In summary, we have updated the POLAR code by including in the modeling of the reionization process the effect of ionization and heating from stellar sources, X-ray binaries and shock heated ISM, as well as of an evolving galactic star formation history.
Since different SFH of galaxies can {affect the recombination and cooling of ionized and heated IGM gas and as a consequence} result in different ionizing and heating patterns, an accurate modeling of the galactic SFH is important in investigations of the 21-cm signal.

\begin{acknowledgements}
This work is supported by the National Natural Science Foundation of China (Grant No. 12263002).
RG acknowledges support from SERB, DST Ramanujan Fellowship no. RJF/2022/000141.
\end{acknowledgements}

\bibliographystyle{aa} 
\bibliography{aa} 

\begin{appendix}
{
\onecolumn
\section{Star formation history with 1-$\sigma$ scatter}
\label{SFH_scatter}
\begin{figure*}
    \centering
    \includegraphics[width=0.95\linewidth]{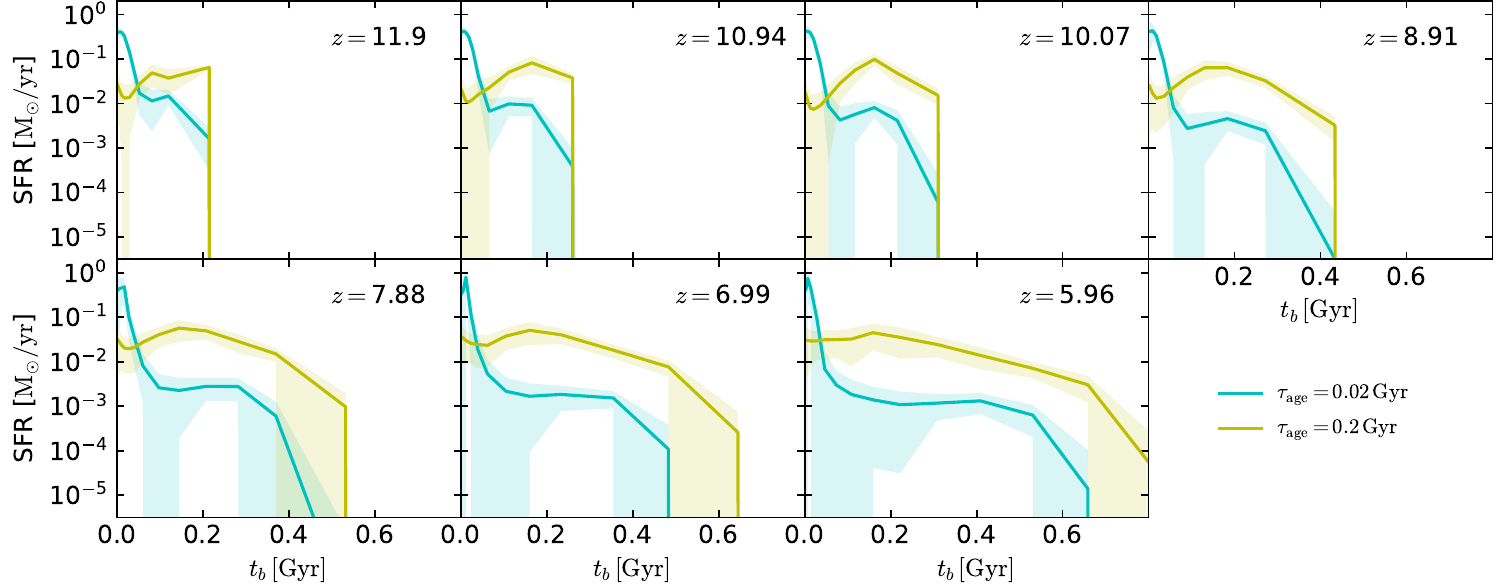}
    \caption{Average SFR history with 1-$\sigma$ scatter of galaxies with $M_{\star} \sim 10^{7}\,\rm M_{\odot}$ and stellar age $\tau_{\rm age}=0.02\,\rm Gyr$ (cyan), and $0.2\,\rm Gyr$ (yellow) at $z=11.9$, 10.94, 10.07, 8.91, 7.88, 6.99 and 5.96 from left to right and top to bottom.
    The x-axis $t_{b}$ is the time of galaxies traced back from $z$ to higher redshift.
    }
    \label{fig:sfh_mulitz_error}
\end{figure*}
Figure~\ref{fig:sfh_mulitz_error} shows the history of the average SFR, together with the 1-$\sigma$ scatter, of galaxies with $M_{\star} \sim 10^{7}\,\rm M_{\odot}$ and different $\tau_{\rm age}$ at seven redshifts in the range 12 to 6.
Since there is a substantial overlap of the 1-$\sigma$ scatter of many curves, here we show the results only for two representative $\tau_{\rm age}$.
We note that the differences in the SFH of galaxies with different $\tau_{\rm age}$ are obvious even when accounting for the scatter.

Since using the SFH of each galaxy in {\sc Grizzly} is too computationally expensive, we will take the average SFH for all the galaxies within a given stellar mass and age bin.
Although the average SFH has some scatter, we find that this is a good approximation and that it makes the running of {\sc Grizzly} efficient.
As a reference, we adopt 29 stellar mass bins and 13 age bins i.e. 377 galaxy bins at each redshift, while the results of LG20 with the merger trees from {\sc Jiutian-300} has e.g. $\sim 10^8$ galaxies at $z=6$.
}
\FloatBarrier
\twocolumn
\end{appendix}

\end{document}